\documentstyle[12pt,fleqn]{article}
\def\thesection{}
\def\re#1{{[\ref{#1}]}}
\begin{document}
\begin{titlepage}
\vspace*{-62pt}
\begin{flushright}
{\footnotesize
SUSSEX--AST 93/8-2 \\
FNAL--PUB--93/258--A\\
astro-ph/9308044\\
(August 1993)}
\end{flushright}
\renewcommand{\thefootnote}{\fnsymbol{footnote}}
\begin{center}
{\Large \bf Reconstructing the inflaton potential---\\ perturbative
reconstruction to second-order}\\

\vspace{0.6cm}
\normalsize

Edmund J.\ Copeland\footnote{Electronic mail: edmundjc@central.sussex.ac.uk}\\
{\em School of Mathematical and Physical Sciences,\\
University of Sussex, Brighton BN1 9QH, U.\ K.}\\

\vspace{0.4cm}

Edward W.\ Kolb\footnote{Electronic mail: rocky@fnas01.fnal.gov}\\
{\em NASA/Fermilab Astrophysics Center\\
Fermi National Accelerator Laboratory, Batavia, IL~~60510, and\\
Department of Astronomy and Astrophysics, Enrico Fermi Institute\\
The University of Chicago, Chicago, IL~~ 60637}\\

\vspace{0.4 cm}

Andrew R.\ Liddle\footnote{Electronic mail: arl@starlink.sussex.ac.uk}\\
{\em Astronomy Centre, School of Mathematical and Physical Sciences,\\
University of Sussex, Brighton BN1 9QH, U.\ K.}\\

\vspace{0.4cm}

James E.\ Lidsey\footnote{Electronic mail: jim@fnas09.fnal.gov}\\
{\em NASA/Fermilab Astrophysics Center\\
Fermi National Accelerator Laboratory, Batavia, IL~~60510}

\end{center}

\vspace*{12pt}

\begin{quote}
\hspace*{2em}One method to reconstruct the scalar field potential of inflation
is a perturbative approach, where the values of the potential and its
derivatives are calculated as an expansion in departures from the slow-roll
approximation. They can then be expressed in terms of observable quantities,
such as the square of the ratio of the gravitational wave amplitude to the
density perturbation amplitude, the deviation of the spectral index from the
Harrison--Zel'dovich value, etc. Here, we calculate complete expressions for
the second-order contributions to the coefficients of the expansion by
including for the first time corrections to the standard expressions for the
perturbation spectra. As well as offering an improved result, these corrections
indicate the expected accuracy of the
reconstruction. Typically the corrections are only a few percent.

\vspace*{12pt}

\small
PACS numbers: 98.80.-k, 98.80.Cq, 12.10.Dm
\end{quote}
\renewcommand{\thefootnote}{\arabic{footnote}}
\addtocounter{footnote}{-4}

\end{titlepage}
\thesection{\centerline{\bf I. INTRODUCTION}}
\setcounter{section}{1}
\setcounter{equation}{0}
\vspace{18pt}

An intriguing prospect raised by recent large-scale structure observations,
particularly those of the Cosmic Background Explorer (COBE) satellite
\re{DMR}, is that  observations may soon provide rather detailed information
regarding the nature of the vacuum energy driving inflation
[\ref{INF},\ref{KT},\ref{LL}].  In most models this vacuum energy is identified
as  the self-interaction potential of a scalar {\em inflaton} field, and its
precise form is determined by some particle physics model. Since there
currently exist many possible models \re{MODELS}, it is of great interest to
investigate whether observations can select which, if any, of these models is
correct. As well as having detailed implications for the initial conditions for
structure formation in the Universe, the energy scale of inflation provides a
link with particle physics at high energies, and may be a useful method of
probing models of unification.

The prime observational consequences of inflation derive from the stochastic
spectra of density (scalar) perturbations and gravitational wave (tensor)
modes generated during inflation. Each stretches from scales of order
centimeters to scales well in excess of the size of the presently observable
Universe. Once within the Hubble radius, gravitational waves redshift away and
so their main influence is on the large-scale microwave background
anisotropies, such as those probed by COBE \re{TENSORS}. Advanced
gravitational wave detectors such as the proposed beam-in-space experiments
may be able to detect the gravitational waves on a much shorter (about
$10^{14}$cm) wavelength range \re{LIDGRAV}. The density perturbations are
thought to lead to structure formation in the Universe. They produce microwave
background anisotropies across a much wider range of angular scales than do the
tensor modes, and constraints on the scalar spectrum are also available from
the clustering of galaxies and galaxy clusters, peculiar velocity flows and a
host of other measurable quantities \re{LL}.

Recently, we provided a formalism which allows one to reconstruct the
inflaton potential $V(\phi)$ directly from a knowledge of these spectra
[\ref{CKLL1},\ref{CKLL2}]. This developed an original but incomplete analysis
by Hodges and Blumenthal \re{HB}. An important result that follows from our
formalism is that knowledge of the scalar spectrum alone is insufficient for a
unique reconstruction.  Reconstruction from only the scalar spectrum
leaves an arbitrary integration constant, and since the reconstruction is
nonlinear, different choices of this constant lead to different functional
forms for the potential. A minimal knowledge of the tensor spectrum, say its
amplitude at a single wavelength, is sufficient to lift this degeneracy. With
further information the problem becomes overdetermined, providing powerful
consistency relations which would exclude inflation if not satisfied.

The most ambitious aim of reconstruction is to employ observational data to
deduce the complete functional form of the inflaton potential over the range
corresponding to large-scale structure. The observational situation is some
way from providing the quality of data that this would require, and at present
a more realistic approach is to attempt a  reconstruction  of the potential
about a single point $\phi_0$ \re{CKLL2}.

For this one requires such information as the amplitudes of both scalar and
tensor modes and
also the spectral index of the scalar perturbations at a single scale. It is
possible that such information can be deduced from a combination of microwave
background experiments that span a range of angular scales \re{CBDES}. In
this paper it is our aim to provide an improved calculation of the
coefficients of such a perturbative reconstruction.

To some extent all inflationary calculations rely on the use of the slow-roll
approximation. In the form we present here, the slow-roll approximation is an
expansion in terms of quantities defined from derivatives of the Hubble
parameter $H$. In general there are an infinite hierarchy of these which
can in principle all enter at the same order in an expansion. However, to
calculate $V(\phi_0)$ one needs only the first derivative of $H$, for
$V'(\phi_0)$ one needs up to the second derivative of $H$, and so on. These
parameters can be converted into more observationally related quantities, as we
shall see.

The slow-roll approximation arises in two separate places. The first is in
simplifying the classical inflationary dynamics of expansion, and the
lowest-order approximation  ignores the contribution of the inflaton's kinetic
energy to the expansion rate. The second is in the calculation of the
perturbation spectra; the standard expressions are true only to lowest-order
in slow-roll. In our earlier work [\ref{CKLL1},\ref{CKLL2}], we utilized the
Hamilton-Jacobi approach \re{HJ} to treat the dynamical evolution exactly,
but were forced for analytic tractability to retain the lowest-order
approximation for the perturbation spectra. Technically therefore, the results
were accurate only to lowest-order, though in models close to the power-law
inflation limit this hybrid approach offers substantial improvements for
certain quantities.

Until recently further improvements have not been possible, but a very elegant
calculation of the perturbation spectra to next order in slow-roll has now been
provided by Stewart and Lyth \re{SL}. This does not permit analytic progress in
functional reconstruction, but their results can be combined with the
Hamilton-Jacobi approach to generate the complete second-order term in
perturbative reconstruction. The purpose of this work is to calculate this
correction.  This serves two useful purposes. Firstly, the results allow a more
accurate reconstruction to be performed, and secondly  the relative size of
lowest-order and second-order contributions provides a useful (though not
rigorous) measure of the theoretical error in reconstruction. As we shall see,
even the lowest-order results are typically accurate to within a few percent.

\vspace{48pt}
\thesection{\centerline{\bf II. TO SECOND-ORDER IN SLOW-ROLL}}
\setcounter{section}{2}
\setcounter{equation}{0}

\vspace{12pt}

We shall  use the notation and philosophy of our earlier paper \re{CKLL2},
except that the perturbation spectra shall be given by expressions improving
on Eq.\ (3.4) of that work. The Hamilton-Jacobi equations arise when one uses
the scalar field $\phi$ as a time variable, and writes the Hubble parameter $H
= \dot{a}/a$, where $a$ is the scale factor, as a function of $\phi$. The
field equations are \re{HJ}
\begin{eqnarray}
\label{eom}
[H'(\phi)]^2 - \frac{3}{2} \kappa^2 H^2(\phi) & = & - \frac{1}{2} \kappa^4
V(\phi) \,, \\
\kappa^2 \dot{\phi} & = & -2 H' \,,
\end{eqnarray}
where dots are time derivatives, primes are $\phi$ derivatives,  $\kappa^2 =
8\pi/m^{2}_{Pl}$ and $m_{Pl}$ is the Planck mass. Without loss of generality
we may assume  $\dot{\phi} > 0$, so that $H'(\phi) < 0$. Where square roots
appear later this choice is used to fix the sign of the prefactor.

The slow-roll approximation can be specified by parameters defined from
derivatives of $H(\phi)$. There are in general an infinite number of these as
each derivative is independent, but usually only the first few enter into any
expressions. We shall require the first three, which are all of the
same order when defined by\footnote{Let us stress that our choice $\dot{\phi} >
0$ implies $\sqrt{\epsilon} = - \sqrt{2/\kappa^2} \, H'/H$; one needs to be
careful with the signs to reproduce our results.}
\begin{eqnarray}
\label{def}
\epsilon(\phi) & = & \frac{2}{\kappa^2} \left[ \frac{H'(\phi)}{H(\phi)}
\right]^2 \,, \nonumber \\
\eta(\phi) & = & \frac{2}{\kappa^2} \, \frac{H''(\phi)}{H(\phi)} = \epsilon
- \frac{\epsilon'}{\sqrt{2\kappa^2 \epsilon}}\,, \nonumber \\
\xi(\phi) & = & \frac{2}{\kappa^2} \, \frac{H'''(\phi)}{H' (\phi)} = \eta -
	\frac{2\eta'}{\sqrt{2\kappa^2 \epsilon}} \,.
\end{eqnarray}
The slow-roll approximation applies when these slow-roll parameters are small
in comparison to unity. The condition for inflation, $\ddot{a} > 0$, is
precisely equivalent to $\epsilon < 1$.

The lowest-order expressions for the scalar $(A_S)$ and tensor $(A_G)$
amplitudes assume $\{\epsilon,\ \eta,\ \xi\}$ are negligible compared to unity.
Improved expressions for the scalar and tensor amplitudes for finite but small
$\{\epsilon,\ \eta,\ \xi\}$ were found by Stewart and Lyth \re{SL}:
\begin{eqnarray}
\label{scalar}
A_S & \simeq & -  \frac{\sqrt{2} \kappa^2}{8\pi^{3/2}} \, \frac{H^2}{H'} \,
\left[ 1 - (2C+1)\epsilon + C \eta \right]\, , \\
\label{wave}
A_G & \simeq & \frac{\kappa}{4\pi^{3/2}} \, H \, \left[ 1 - (C+1) \epsilon
\right] \,  ,
\end{eqnarray}
where $C = -2 + \ln 2 + \gamma \simeq -0.73$ is a numerical constant, $\gamma
\approx 0.577$ being the Euler constant. The right hand sides of these
expressions are evaluated when the scale in question crosses the Hubble radius
during inflation, $2\pi/\lambda = aH$. The spectra can equally well be
considered to be functions of wavelength or of the scalar field value. Eq.\
(\ref{dlambda}) below allows one to move from one to the other.

The standard results to lowest-order are given by setting the square brackets
to unity. Historically it has been common even for this result to be written
as only an approximate equality (the ambiguity arising primarily because of a
vagueness in defining the precise meaning of the density perturbation), though
the precise normalization to lowest-order was established some time ago by Lyth
\re{L85} (see also the discussion in \re{LL}).

The improved expressions for the spectra in Eqs.\ (\ref{scalar}) and
(\ref{wave}) are accurate in so far as $\epsilon$ and $\eta$ are sufficiently
slowly varying functions that they can be treated adiabatically as constants
while a given scale crosses outside the Hubble radius. Corrections to this
would enter at next order. This differs from  the usual situation in which $H$
is treated adiabatically. For the standard calculation to be strictly valid $H$
must be constant, but provided it varies sufficiently slowly (characterized by
small $\epsilon$ and $|\eta|$), it can be evaluated separately at each epoch.
This  injects a scale dependence into the spectra. There is a special case
corresponding to power-law inflation for which $\epsilon$ and $\eta$ are
precisely constant and equal to each other. In this case the above expressions
for the perturbation spectra are exact [\ref{SL},\ref{LS}]. Furthermore, the
corrections to each spectrum are the same and they cancel when the ratio is
taken. In the general case $\epsilon$ and $\eta$ may be treated as different
constants if it is assumed that the timescale for their evolution is much
longer than the timescale for perturbations to be imprinted on a given scale.
This assumption worsens as $\eta$ is removed from $\epsilon$, which would be
characterized by the next order terms becoming large.

Throughout we shall be quoting results which feature a leading term and a
correction term linear in the slow-roll parameters. We shall utilize the
symbol ``$\simeq$'' to indicate this level of accuracy throughout. The
correction terms shall be placed in square brackets, so the lowest-order
equations can always be obtained by setting the square brackets equal to one. A
useful relationship can be obtained from Eqs.\ (\ref{def})--(\ref{wave}):
\begin{equation}
\epsilon \simeq \frac{A_G^2}{A_S^2} \left[ 1 - 2 C (\epsilon -  \eta ) \right]
\,.
\end{equation}
As we shall see, $\eta$ is encoded in the spectral index of the scalar
perturbations, whose deviation from unity must also be small for slow-roll to
apply.

A key equation in Refs.\ [\ref{CKLL1},\ref{CKLL2}] is the consistency equation,
which connects the scalar spectrum to the tensor spectrum and its derivative.
The spectra as given in Eqs.\ (\ref{scalar}) and (\ref{wave}) are  functions of
the value of $\phi$ when the fluctuations crossed the Hubble radius during
inflation. This is converted into a dependence on wavelength $\lambda$ with the
relation \re{CKLL2}
\begin{equation}
\label{dlambda}
\frac{d\lambda}{d\phi} = \lambda \frac{H}{H'} \frac{\kappa^2}{2}
	\left[ 1 - \epsilon \right] \,.
\end{equation}
Differentiation of Eq.\ (\ref{wave}) with respect to $\phi$ implies that
\begin{equation}
\frac{d\ln A_G}{d\phi} \simeq - \sqrt{\frac{\kappa^2}{2}} \,
	\frac{A_G}{A_S} \left[ 1 + (C+2) \epsilon - (C+2) \eta \right] \,,
\end{equation}
and it follows that
\begin{equation}
\label{consistency}
\frac{\lambda}{A_G} \frac{dA_G}{d\lambda} = \frac{A_G^2}{A_S^2}
	\left[ 1 + 3 \epsilon - 2 \eta \right] \,.
\end{equation}
As expected this agrees  with the expansion of the corresponding expression
in Ref.\ \re{CKLL2} for the special case $\epsilon = \eta$. This equation is
interesting in its own right, but for perturbative reconstruction its use is
restricted to the removal of derivatives of the tensor spectrum.

\vspace{48pt}
\thesection{\centerline{\bf III. PERTURBATIVE RECONSTRUCTION TO SECOND ORDER}}
\setcounter{section}{3}
\setcounter{equation}{0}

\vspace{12pt}

The aim now is to obtain expressions for the potential and its derivatives
about a single point $\phi_0$, given information regarding the spectra at the
scale $\lambda_0$ which left the horizon at $\phi=\phi_0$. The four main
quantities of observational interest are the amplitudes and spectral indices
of the two spectra. However, in view of the consistency  equation, Eq.\
(\ref{consistency}), only three of these are independent. We shall concentrate
on the two amplitudes and the scalar spectral index, since these are probably
the easiest to measure. The scalar spectral index $n$ is defined by
\begin{equation}
1-n = \frac{d\ln A_S^2(\lambda)}{d \ln\lambda} \,.
\end{equation}
To lowest-order in slow-roll one can show that the spectral index is given by
\re{CKLL2}
\begin{equation}
1-n = 4\epsilon - 2\eta \,,
\end{equation}
which provides the route to determining $\eta$. Conceptually we are passing
from these three observables to the three parameters that describe the
potential, which are the slow-roll parameters $\epsilon$ and $\eta$, and the
overall normalization. In terms of the observables, therefore, the slow-roll
approximation amounts to an expansion in both $A_G^2/A_S^2$ and $(1-n)$, which
give corrections to the same order. We shall see that the correction term for
$V''(\phi_0)$ requires the introduction of the third slow-roll parameter $\xi$,
requiring a new independent observable to determine it.

One obtains directly from the field equation Eq.\ (\ref{eom}) and the
definitions of the spectra in Eqs.\ (\ref{scalar}) and (\ref{wave}) an
expression for the amplitude of the potential:
\begin{eqnarray}
V(\phi_0) & \simeq & \frac{48 \pi^3}{\kappa^4} A_G^2(\lambda_0) \left[ 1 +
	\left( \frac{5}{3} + 2C \right)
	\frac{A_G^2(\lambda_0)}{A_S^2(\lambda_0)} \right] \,, \\
 & \simeq & \frac{48 \pi^3}{\kappa^4} A_G^2(\lambda_0) \left[ 1 +
	0.21 \frac{A_G^2(\lambda_0)}{A_S^2(\lambda_0)}\right] \,.
\end{eqnarray}
In Refs.\ [\ref{CKLL1},\ref{CKLL2}], we gave the numerical factor on the
second-order term as $-1/3$, which incorporated only the dynamical slow-roll
corrections. In fact, the spectral corrections to $V(\phi_0)$ dominate the
dynamical ones for any inflation model, reversing the sign of the correction,
which may be significant if the tensors are important. However the relative
contribution   of the scalar and tensor modes to large angle microwave
anisotropies with our spectral normalization is given approximately by $R$
\re{CKLL2}, where
\begin{equation}
R=\frac{2  A_S^2}{25 A_G^2} \,.
\end{equation}
Even if the contributions to the anisotropies from scalar and tensor modes are
equal,  the correction term in the potential is only 2\%. This is a powerful
indication that even the lowest-order perturbative reconstruction promises to
be very accurate.

To obtain $V'(\phi_0)$ we need the scalar spectral index at $\lambda_0$,
denoted by $n_0$. Differentiation of the potential with respect to $\phi$,
followed by some straightforward algebra gives
\begin{eqnarray}
V'(\phi_0) & \simeq & - \frac{96 \pi^3}{\sqrt{2} \kappa^3} \,
	\frac{A_G^3(\lambda_0)}{A_S(\lambda_0)} \left[ 1 + (C+2) \epsilon +
	(C-1/3) \eta \right] \,, \nonumber \\
 & \simeq & - \frac{96 \pi^3}{\sqrt{2} \kappa^3} \,
	\frac{A_G^3(\lambda_0)}{A_S(\lambda_0)} \left[ 1 + 1.27 \epsilon
	-1.06 \eta \right] \,, \nonumber \\
 & \simeq & - \frac{96 \pi^3}{\sqrt{2} \kappa^3} \,
	\frac{A_G^3(\lambda_0)}{A_S(\lambda_0)} \left[ 1 -0.85
	\frac{A_G^2(\lambda_0)}{A_S^2(\lambda_0)} +0.53 (1-n_0) \right] \,.
\end{eqnarray}
Note that for power-law inflation, which has \re{TENSORS}
\begin{equation}
(1-n_0) \simeq \frac{25}{4\pi} \, \frac{A_G^2(\lambda_0)}{A_S^2(\lambda_0)} \,,
\end{equation}
the corrections in the square brackets nearly cancel, but other models
[\ref{NAT},\ref{JIML}] can feature larger corrections, {\it e.g.} 16 \% for
natural inflation with $n_0 = 0.7$.

The calculation for $V''(\phi_0)$ is much more involved. One can show a
precise relationship
\begin{equation}
\label{+++++++}
\frac{V''(\phi_0)}{H^2(\phi_0)} = 3 (\epsilon + \eta) - \left(\eta^2 +
	\epsilon \xi \right) \,.
\end{equation}
A new observable will be needed to determine $\xi$, the easiest example being
the rate of change of the scalar spectral index. This would be substantially
harder to measure, and it is fortunate that it only enters at second-order.
[It would however enter at leading order in $V'''(\phi_0)$]. From Eqs.
(\ref{wave}) and (\ref{+++++++}),
we can obtain the second-order correction to $V''(\phi_0)$ in terms of the
slow-roll parameters
\begin{equation}
\label{VPPSR}
V''(\phi_0) \simeq \frac{48 \pi^3}{\kappa^2} A_G^2(\lambda_0) \left( \epsilon
	+ \eta \right) \left[ 1 + (2C+2) \epsilon -
	\frac{\eta^2 + \epsilon \xi}{3(\epsilon+\eta)} \right] \,.
\end{equation}
It is however   harder to convert the prefactor into observables,
because there are now lowest-order terms in both $\epsilon$ and $\eta$. To
generate the correct second-order term, it is not enough to use the
first-order expression for $\eta$ in terms of the spectral index. One must
instead use the second-order result, as given by Stewart and Lyth \re{SL}
\begin{equation}
1-n \simeq 4 \epsilon - 2 \eta + 8(1+C) \epsilon^2 - (6+10C) \epsilon \eta
	+ 2C \epsilon \xi \,,
\end{equation}
(our $\eta$ being the negative of their $\delta$), which leads to
\begin{eqnarray}
\epsilon+\eta & \simeq & 3 \epsilon \left[ 1 + \frac{4C+4}{3} \epsilon
	- \frac{5C+3}{3} \eta + \frac{C}{3} \xi \right] -
	\frac{1-n_0}{2} \,, \nonumber \\
 & \simeq & 3 \frac{A_G^2}{A_S^2} \left[ 1 + \frac{4-2C}{3} \epsilon +
	\frac{C-3}{3} \eta + \frac{C}{3} \xi \right] - \frac{1-n_0}{2} \,.
\end{eqnarray}
Substituting this into Eq.\ (\ref{VPPSR}) yields
\begin{eqnarray}
\label{***}
V''(\phi_0) & \simeq & \frac{144\pi^3}{\kappa^2}
	\frac{A_G^4(\lambda_0)}{A_S^2(\lambda_0)} \left[ 1 + \frac{4C+10}{3}
	\epsilon + \frac{C-3}{3} \eta + \frac{C}{3} \xi \right] \\
 & & - \frac{24 \pi^3}{\kappa^2} A_G^2(\lambda_0) \left(1-n_0 \right) \left[
	1 + (2C+2) \epsilon \right] - \frac{16\pi^3}{\kappa^2} A_G^2
	(\eta^2+\epsilon \xi) \,,  \nonumber
\end{eqnarray}
where the last term is entirely second-order. Note that there are two
lowest-order terms. An interesting case is  $\eta= -\epsilon$, corresponding to
$H \propto \phi^{1/2}$, for which the lowest-order term vanishes identically
and the final term of Eq.\ (\ref{VPPSR}) is the only one to contribute. The
second derivative of the potential is the lowest derivative at which it is
possible for the expected lowest-order term to vanish.

The final step is to convert the second-order terms into the observables. As
they are already second-order, one only needs the lowest term in their
expansion to convert. From the expression for the spectral index, one finds to
lowest-order that
\begin{equation}
\xi \simeq \frac{1}{2\epsilon} \left. \frac{dn}{d\ln \lambda}
	\right|_{\lambda_0} + 5 \eta - 4 \epsilon \,.
\end{equation}
Note that the derivative of the spectral index is of order $\epsilon^2$. To
lowest-order we also have
\begin{equation}
\eta \simeq 2 \epsilon - \frac{1}{2} (1-n_0);\qquad
\epsilon \simeq \frac{A_G^2}{A_S^2} \,.
\end{equation}
Progressively substituting all these into Eq.\ (\ref{***}) yields
\begin{eqnarray}
V''(\phi_0) & \simeq & \frac{16\pi^3}{\kappa^2} A_G^2(\lambda_0) \left[ 9
	\frac{A_G^2(\lambda_0)}{A_S^2(\lambda_0)} - \frac{3}{2} (1-n_0)
	+ (36C+2) \frac{A_G^4(\lambda_0)}{A_S^4(\lambda_0)}
	\right. \nonumber \\
 & & \left. - \frac{1}{4} (1-n_0)^2 - (12C-6)
	\frac{A_G^2(\lambda_0)}{A_S^2(\lambda_0)} (1-n_0)  \right. \nonumber \\
 & & + \left. \frac{3C-1}{2}
	\left. \frac{dn}{d \ln \lambda} \right|_{\lambda_0}
	\right] ,
\end{eqnarray}
where the first two terms are lowest-order and the remainder are
second-order.\footnote{Factoring out the lowest-order terms as in previous
expressions leads to a very complicated result, so we break with our convention
regarding the use of square brackets.}

For power-law models, the last term is zero and the remaining correction terms
nearly cancel each other, though they are not small individually. For natural
inflation models the correction terms are all individually small, with an
overall correction of about 4\% at $n_0 = 0.7$. (In natural inflation,
$(dn/d\ln \lambda) |_{\lambda_0} \simeq \eta^2 \simeq (1-n_0)^2/16$
\re{SL}.)

\vspace{48pt}
\thesection{\centerline{\bf IV. DISCUSSION AND CONCLUSIONS}}
\setcounter{section}{4}
\setcounter{equation}{0}

\vspace{12pt}

To conclude, we have calculated the full second-order corrections to the
perturbative reconstruction of the inflaton potential. The first-order terms
agree with those we found previously [\ref{CKLL1},\ref{CKLL2}], while the
second-order terms offer an improvement. They serve to quantify the expected
errors in the perturbative reconstruction, and in general these errors are
small. Even in cases where tensors provide a substantial contribution to the
large angle microwave background anisotropies and/or the spectral index
deviates significantly from unity, the corrections are typically only a few
percent.
Consequently, example figures based on plausible data sets that we presented
in our earlier papers remain valid. One then has some degree of confidence that
one can use our lowest-order expressions as was done in \re{MIKE}.

\vspace{18pt}

\centerline{\bf ACKNOWLEDGMENTS}

EJC, ARL and JEL are supported by the Science and Engineering Research Council
(SERC) UK. EWK and JEL are supported at Fermilab by the DOE and NASA under
Grant NAGW--2381. ARL acknowledges the use of the Starlink computer system at
the University of Sussex. EJC and ARL thank the Aspen Center for Physics for
hospitality during the completion of this work. We would like to thank David
Lyth for helpful discussions.

\frenchspacing
\def\prl#1#2#3{Phys. Rev. Lett. {\bf #1}, #2 (#3)}
\def\prd#1#2#3{Phys. Rev. D{\bf #1}, #2 (#3)}
\def\plb#1#2#3{Phys. Lett. {\bf #1B}, #2 (#3)}
\def\npb#1#2#3{Nucl. Phys. {\bf B#1}, #2 (#3)}
\def\apj#1#2#3{Astrophys. J. {\bf #1}, #2 (#3)}
\def\apjl#1#2#3{Astrophys. J. Lett. {\bf #1}, #2 (#3)}

\begin{picture}(400,50)(0,0)
\put (50,0){\line(350,0){300}}
\end{picture}

\vspace{0.25in}

\def\labelenumi{[\theenumi]}

\begin{enumerate}
\item \label{DMR} G. F. Smoot {\it et al.},  \apjl{396}{L1}{1992};
	E. L. Wright {\it et al.}, \apjl{396}{L13}{1992}.
\item \label{INF} A. Guth, \prd{23}{347}{1981}; A. Albrecht and P. J.
	Steinhardt, \prl{48}{1220}{1982}; A. D. Linde, \plb{108}{389}{1982};
	A. D. Linde, \plb{129}{177}{1983}.
\item \label{KT} E. W. Kolb and M. S. Turner, {\em The Early Universe},
	(Addison-Wesley, New York, 1990).
\item \label{LL} A. R. Liddle and D. H. Lyth, ``The Cold Dark Matter
	Density Perturbation,'' to be published, Phys. Rep. (1993).
\item \label{MODELS} See Refs.\ \ref{KT} and \ref{LL} for surveys of
	different models.
\item \label{TENSORS} L. M. Krauss and M. White, \prl{69}{869}{1992}; R. L.
	Davis, H. M. Hodges, G. F. Smoot, P. J. Steinhardt and M. S. Turner,
	\prl{69}{1856}{1992}; A. R. Liddle and D. H. Lyth,
	\plb{291}{391}{1992}; J. E. Lidsey and P. Coles, Mon. Not. R. astr.
	Soc. {\bf 258}, 57P (1992); D. S. Salopek, \prl{69}{3602}{1992};
	F. Lucchin, S. Matarrese, and S. Mollerach, \apjl{401}{49}{1992}.
\item \label{LIDGRAV} A. R. Liddle, ``Can the Gravitational Wave Background
	from Inflation be Detected Locally?'' Sussex report SUSSEX-AST
	93/7-3 (1993).
\item \label{CKLL1} E. J. Copeland, E. W. Kolb, A. R. Liddle and J. E.
	Lidsey, \prl{71}{219}{1993}.
\item \label{CKLL2} E. J. Copeland, E. W. Kolb, A. R. Liddle and J. E.
	Lidsey, to appear, Phys. Rev. D,  (Sept 15th 1993).
\item \label{HB} H. M. Hodges and G. R. Blumenthal, \prd{42}{3329}{1990}.
\item \label{CBDES} R. Crittenden, J. R. Bond, R. L. Davis, G. P. Efstathiou
	and P. J. Steinhardt, \prl{71}{324}{1993}.
\item \label{HJ} D. S. Salopek and J. R. Bond, \prd{42}{3936}{1990};
	J. E. Lidsey, \plb{273}{42}{1991}.
\item \label{SL} E. D. Stewart and D. H. Lyth, \plb{302}{171}{1993}.
\item \label{L85} D. H. Lyth, \prd{31}{1792}{1985}.
\item \label{LS} D. H. Lyth and E. D. Stewart, \plb{274}{168}{1992}.
\item \label{NAT} F. C. Adams, J. R. Bond, K. Freese, J. A. Frieman and A. V.
	Olinto, \prd{47}{426}{1993}.
\item \label{JIML} J. E. Lidsey, ``Tilting the Primordial Power Spectrum with
	Bulk Viscosity,'' Fermilab report 93/099--A (1993).
\item \label{MIKE} M. S. Turner, ``Recovering the Inflationary Potential,''
 	Fermilab report 93/182-A (1993).
\end{enumerate}
\end{document}